\documentclass[prl,twocolumn,showpacs,amsmath,amssymb,superscriptaddress]{revtex4}

\usepackage{graphicx}
\usepackage{color}
\usepackage{pslatex}

\newcommand{\void}[1]{}
\renewcommand{\nu}{2\epsilon-\Delta} 

\begin{document}

\title{Non-equilibrium Effects in the Thermal Switching of 
  Underdamped Josephson Junctions}

\author{Juan Jos\'e Mazo}
\address{Departamento de F\'{\i}sica de la Materia Condensada
and Instituto de Ciencia de Materiales de
  Arag\'on, CSIC-Universidad de Zaragoza, 50009 Zaragoza, Spain}

\author{Fernando Naranjo} \address{Departamento de F\'{\i}sica de la
  Materia Condensada and Instituto de Ciencia de Materiales de
  Arag\'on, CSIC-Universidad de Zaragoza, 50009 Zaragoza, Spain}
\address{Universidad Pedag\'ogica y
  Tecnol\'ogica de Colombia, Tunja, Colombia}

\author{David Zueco}
\address{Departamento de F\'{\i}sica de la Materia Condensada
and Instituto de Ciencia de Materiales de
  Arag\'on, CSIC-Universidad de Zaragoza, 50009 Zaragoza, Spain}

\date{\today}

\begin{abstract}
We study the thermal escape problem in the low damping limit. We find
that finiteness of the barrier is crucial for explaining the thermal
activation results. In this regime low barrier non-equilibrium
corrections to the {\it usual} theories become necessary.  We propose
a simple theoretical extension accounting for these non-equilibrium
processes which agrees numerical results.  We apply our theory to the
understanding of switching current curves in underdamped Josephson
junctions.

\end{abstract}

\maketitle

In 1940 Kramers derived his famous formulas describing rates in
chemical reactions~\cite{Kramers1940}. The theoretical framework for
his calculation was the escape of a Brownian particle over a potential
barrier. Far from being a particular case,
noise-activated escape is applicable in a wide number of problems in
science, going from biology to quantum information
processing~\cite{Pollak2005a}.  Due to the many fields involved,
intense activity emerged in the subject proposing better theories for
this, nowadays, old problem~\cite{Pollak2005a, Hanggi1990a,
  Melnikov1991a}.

In particular, studies on thermal switching in Josephson junctions
(JJ) benefits from this
effort~\cite{Ambegaokar1969,Stephen1969,Kurkijavi1972,Fulton1974,Washburn1985,Devoret1985,Martinis1987,Silvestrini1988}. Experimental
results are affected by thermal fluctuations and measurements in the
lab allow to predict junctions parameters by fitting the switching
with available expressions. Also, some fundamental issues as the
quantum-classical transition have been addressed by means of rates
measurements
~\cite{Washburn1985,Devoret1985,Martinis1987,Wallraff2003a}.  It is
clear, that such a measurements need to be compared with appropriate
theoretical results. Needless to say the {\it exact} formula does not
exist and many theories are available in the literature, which
starting from the Kramers seminal work, cover different set of
parameters~\cite{Hanggi1990a,Melnikov1991a}.

In a recent numerical work~\cite{Mazo2008}, for very low values of the
damping parameter, it has been found a significant deviation of the JJ
switching current from the expected result. Here we present a theory
that is able to give account for the observed deviation. Moreover, we
predict that this effect will appear in any biased system where the
damping over force ramp ratio is not large. In such a case, the usual
theories are not suitable, and, as we show, it is needed to include
non-equilibrium effects and finite barrier correction in a full
description of the problem.

To be definite, the dynamics for the phase difference in the junction
is usually described by the, so called, resistively and capacitively
shunted junction (RCSJ) model, which is equivalent to the more general
problem of a Brownian particle in a metastable potential:
\begin{equation}
m\ddot{x}+  m \gamma \dot{x}=-\frac{dV}{d x}+\xi(t),
\label{langevin}
\end{equation}
where the potential $V(x)=V_0 (1-\cos{x})-I x$
and $\xi(t)$ is the stochastic force describing
the thermal fluctuations.  Here we consider white thermal noise,
$\langle \xi(t) \rangle = 0$ and $\langle\xi(t)\xi(t')\rangle= 2 m
\gamma k_B T \delta(t-t')$.

For moderate to low values of the damping parameter there exists a
temperature dependent critical current (force) for the system to
switch from a superconducting or locked state ($\langle \dot{x}
\rangle$=0) to a resistive or running one ($m \gamma \langle \dot{x}
\rangle = I$). Such a situation corresponds to the problem of escape
from a metastable well. In switching current experiments many
current-voltage (force-velocity) curves are performed to obtain the
switching current probability distribution function, $P(I)$. The
measured $P(I)$ can be directly related to the thermal activation
rate~\cite{Fulton1974} and experimental results can be understood in
terms of such parameter.

For very weak damping, the Kramers result for the activation rate is
$r_{_{KLD}} = (\gamma J_b/k_{\rm B}T)(\omega_a/2\pi){\rm e}^{- \Delta
  U/k_B T}$. There we recognize the transition-state-theory result
multiplied by a prefactor valid in the very low damping regime. For
our system the action at the barrier $J_b$ is usually approached by
the cubic potential result $J_b=7.2 \Delta U / \omega_a$. Then:
\begin{equation}
\label{Kramers}
r_{_{KLD}} \simeq \frac{7.2 \gamma}{2 \pi} \; \frac{ \Delta U}{k_{\rm B}
  T} {\rm e}^{- \Delta U/k_B T}.
\end{equation}
This equation shows that the rate scales linearly with the damping and
depends only on the barrier height over thermal energy factor, $\Delta
U/k_B T$. This expression is only valid in the low damping and
infinite barrier limit ($\gamma J_b / k_{\rm B}T \ll 1$ and $\Delta U/
k_{\rm B}T \gg 1$).

Many theories have extended the Kramers result to the
moderate-to-small damping regime~\cite{Buttiker1983a, Grabert1988a,
  Pollak1989a, Melnikov1986a, Hanggi1990a} following the infinite
barrier approximation. Given its simplicity, the result of B\"uttiker
{\it et al.}  (BHL)~\cite{Buttiker1983a} has been usually applied in
the JJ literature. Finite barrier corrections have been studied
in~\cite{Melnikov1993a, Ferrando1995a}. More recently, Drozdov and
Hayashi (DH) proposed a new theory which is
not perturbative in the barrier height~\cite{Drozdov1999}.

We are interested in the dynamics of the system in the low damping
limit. In this limit the coupling to the bath is very weak and the
time to reach thermal equilibrium very long ($\sim 1/\gamma$). This
fact has important consequences: For biased systems, escape occurs at
very low values of the $\Delta U/k_BT$ ratio and junctions may escape
before thermal equilibrium is established and thus non-equilibrium
effects dominate the process. In order to study such effects we need
first to know the importance of finite barrier effects in particle
activation problem at low damping and take into account the average
energy of junctions before each switching event.

\textit{Escape at small barrier}--- We will show here that small
barrier effects are very important in the low damping case, the
convergence to the infinite barrier result is very slow and the DH
theory reproduces the numerical results at any barrier.

\begin{figure}[]
    \centering{\includegraphics[width=0.47\textwidth]{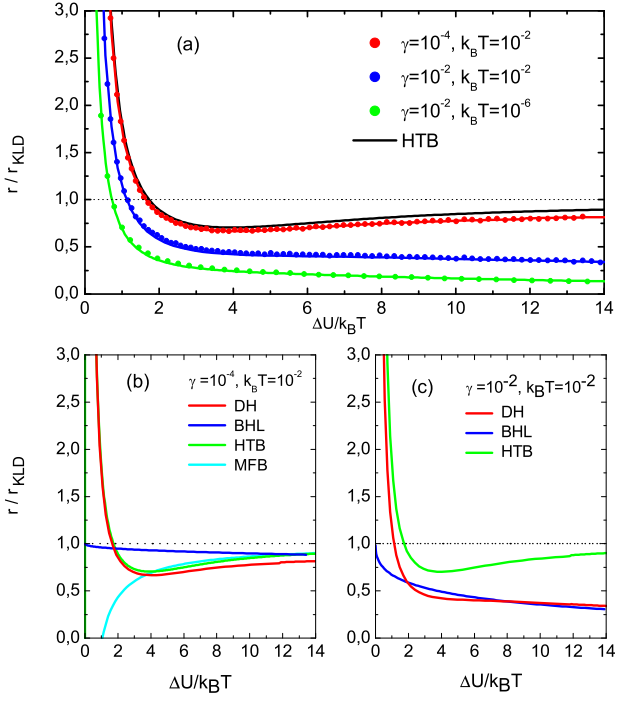}}
    \caption{(color online) Escape rate divided by the Kramers low
      damping result, Eq.~\ref{Kramers}, as a function of $\Delta U
      /k_{\rm B}T$. Figure (a): Dots are numerical results and the
      corresponding color solid lines are the theoretical prediction
      by DH theory~\cite{comm3}. The black line stands for the
      vanishing damping rate formula, Eq. (\ref{HTB}). Figures (b) and
      (c) compare different theoretical results.}
  \label{fig:tasas_barrier}
\end{figure}

\begin{figure}[]
    \centering{\includegraphics[width=0.47\textwidth]{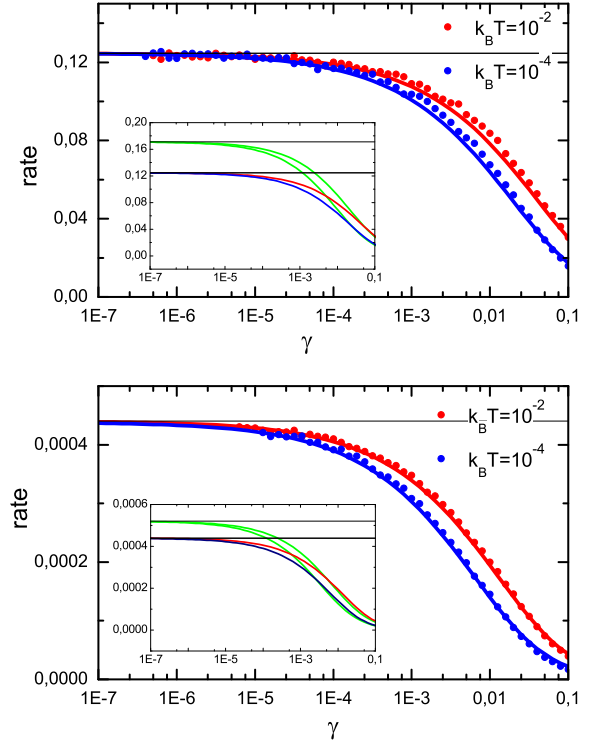}}
    \caption{(color online) Damping dependence for the escape rates.
      The points are numerical results and the corresponding solid
      lines account for DH theory~\cite{comm3}.  The horizontal
      black line is the vanishing damping limit Eq. (\ref{HTB}).  Top:
      curves for $\Delta U /k_{\rm B}T=3$. Bottom: $\Delta U /k_{\rm
        B}T=10$. The inset compares DH to BHL (green lines)
      results.}
  \label{fig:tasas_damping}
\end{figure}

We have numerically integrated the Langevin equation (\ref{langevin})
of the system for different values of damping and barrier height. In
our simulations we have computed the mean time for the system to first
reach the potential barrier. For low values of the damping such mean
time (the first-passage-time problem) corresponds to the inverse of
the escape rate. According to theory, simulations are started with
particles placed in the metastable potential well and zero
velocity. Some issues about the initial conditions problem will be
addressed below. At any point the numerical result is obtained from
$10^4$ escaping events. We show results for $V_0=0.31$, $m=0.35$ and
different values of $F$, damping and temperature.  The results are
summarized in Figs.~\ref{fig:tasas_barrier} and
\ref{fig:tasas_damping} where we plot the activation rate as a
function of barrier and damping respectively and compare to some
existing theories.

Figure~\ref{fig:tasas_barrier} shows the rate dependence on the
barrier for different values of damping and temperature. In order to
see deviations from the Kramers low damping result we divide the
obtained rates by Eq. (\ref{Kramers}).  We recall that (\ref{Kramers})
is obtained assuming weak damping and high barrier. For comparison, we
also plot the exact result for arbitrary barrier in the limit of
vanishing damping~\cite{Hanggi1990a}, $r(\gamma \to 0)=r_{HTB}$ with
\begin{equation}
r_{HTB}= \gamma k_B T \left [ \int_0^{J_b} dJ \ {\rm e}^{-\beta E(J)} 
\int_J^{J_b} dJ' \ \frac{\omega(J')}{2\pi} \ \frac{{\rm e}^{\beta E(J')}}{J'} 
\right ]^{-1},
\label{HTB}
\end{equation}
where $J_b$ is the action at the barrier and $\beta=1/k_BT$.

Remarkably, the approach between both results is slow, meaning that the
high barrier approximation is accurate only at very high barrier
values indeed [Fig.~\ref{fig:tasas_barrier}(a)]. For instance
$r_{HTB}/r_{KLD}=0.72$ for $\Delta U / k_B T=5 $ and 0.85 for $\Delta
U / k_B T= 10$. As a consequence, all theories which try to extend the
Kramers result to the moderate-to-small damping
region~\cite{Buttiker1983a, Grabert1988a, Pollak1989a, Melnikov1986a,
  Hanggi1990a} also fail at low damping values unless very large
barriers are considered.

We know about two main attempts to include finite barrier effects in
this limit: the first one is due to Melnikov
(MFB)~\cite{Melnikov1993a} and fails at small barriers, see
Fig.~\ref{fig:tasas_barrier}(b). The second one was proposed by
Drozdov and Hayashi (DH) for moderate-to-small damping and arbitrary
barrier~\cite{Drozdov1999,comm3}. As it can be seen in
Figs.~\ref{fig:tasas_barrier} and~\ref{fig:tasas_damping}, the DH
theory recovers the vanishing damping limit and explain our numerical
results in the whole small damping region. In
Figs~\ref{fig:tasas_barrier}(b) and (c) we show BHL result for
$\alpha=1$. Other extensions of the Kramers low damping result to the
moderate-to-low damping regime give a result quite similar to BHL.

In Fig.~\ref{fig:tasas_damping} we plot the rate damping dependence
for two different barriers.  Apart from the agreement to the DH
theory, we check the convergence to the very weak limit given by
(\ref{HTB}) [Cf. horizontal line in both figures]. In the inset we
also show results for the BHL.  We see that the finite barrier
corrections become less important by increasing the damping.

To complete our discussion we notice that by decreasing the damping
the curves at different temperatures become the same, that is the rate
depends only on the ratio $\Delta U / k_{\rm B}T$.  This can be
understood by noticing that most of the contribution in the integrals
in Eq.  (\ref{HTB}) comes from the bottom of the potential.  If the
action inside the well is approximated by the corresponding one for an
harmonic potential with the same bottom frequency, $E=(\omega_0/2 \pi)
J $, with $\omega_0 = \partial^2_x V(x_{min})$, the rate (\ref{HTB})
can be written as,
\begin{equation}
\label{HTB-harmonic}
r_{HA} = \gamma \left[ \int^{\Delta U /k_{\rm B}T}_0 \; {\rm d}x \; {\rm e}^{-x} \;
\int^{\Delta U /k_{\rm B}T}_x \; {\rm d}y \; \frac{{\rm e}^{\ y}}{y}
\right]^{-1},
\end{equation}
which clearly depends only on the ratio $\Delta U / k_{\rm B}T$. Let
us emphasize that, besides its simplicity, the above equation is an
excellent approximation to (\ref{HTB}). In fact, plotting both
(\ref{HTB-harmonic}) and (\ref{HTB}) in Fig.~\ref{fig:tasas_barrier},
they cannot be distinguished one from the other.

\begin{figure}[t]
\centering{
          \includegraphics[width=0.47\textwidth]{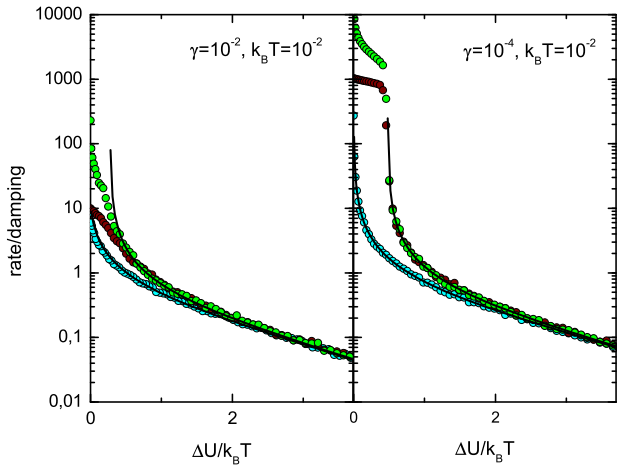}}
            \caption{(color online) Activation rate vs barrier for
              three initial conditions: Particles are at the bottom of
              the metastable well with zero velocity (blue points),
              $v=\sqrt{k_{\rm B}T /m}$ (green points) or
              $v=-\sqrt{k_{\rm B}T /m}$ (brown points).  Lines
              stand for theoretical predictions of Eq. (\ref{mnz}).}
  \label{fig:ci}
\end{figure}

\textit{Initial Conditions dependence}---
We address now the influence of the initial condition for
the energy on the escape rate results. Thermal escape at low damping
is an energy diffusion problem. Escape occurs as soon as thermal
fluctuations provide a particle energy enough to overcome the
barrier. This time depends on the value of the particle initial energy.

Up to know, to compare our simulations to theory we assumed that the
particle starts at the bottom of the metastable well with zero
velocity.  From the experimental point of view this assumption may
fail. Thermal fluctuations not only provide energy enough to surmount
the barrier but also kinetic energy at the bottom. In order to study
the importance of this issue, in Fig.~\ref{fig:ci} we plot the rates
with two initial conditions, $v= \pm \sqrt{k_{\rm B}T /m}$ and compare
to the one with zero velocity. As expected, we see that for small
barriers initial kinetic energy speed up the escape times.

When particles are placed with zero velocity at the bottom of the
well, the activation time is 
$r^{-1}$. However, if particles have extra initial energy $E_{in}$ the escape
time is given by $r^{-1} - \tau$ where $\tau$ is the activation time
up to this extra energy, which can be computed at low damping from
Eq.(\ref{HTB-harmonic}) replacing $\Delta U$ by $E_{in}$.  Putting all
together we generalize the rate formulas as,
\begin{equation}
\label{mnz}
r_{in} = \frac{1}{r^{-1} - \tau}.
\end{equation}
This equation shows that as soon as $\tau(E_{in}) \sim r^{-1}$ the
initial conditions problem affect the escape rates.  In
Fig.~\ref{fig:ci}, where $E_{in}=k_{\rm B} T/2$, this correction
becomes important for $\Delta U / k_{\rm B} T \lesssim 2$. If $\Delta U
\leq E_{in}$ the passage time is almost a deterministic process which
depends on the initial position and velocity of the
system. Figure~\ref{fig:ci} illustrates this effect and confirms our
theoretical prediction.

\begin{figure}[t]
\centering{ \includegraphics[width=0.48\textwidth]{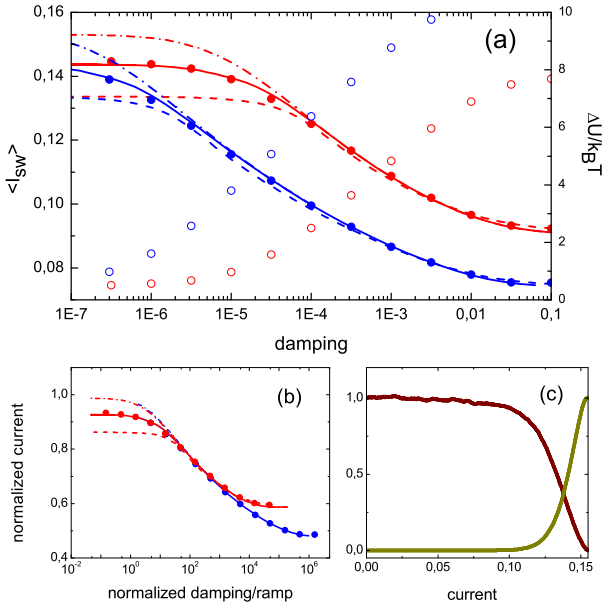}}
            \caption{(color online) (a) Average switching current at
              different values of the damping for $T=0.01$ and current
              ramp $\dot{I}=3.33 \times 10^{-7}$ (red) and
              $\dot{I}=10^{-8}$ (blue). Solid symbols are for
              numerical simulations, dashed lines for predictions
              based on BHL theory, dotted lines for DH theory and
              solid lines for our theory, Eq.~(\ref{mnz}) with
              $E_{in}=k_BT/2$. We also plot (right axis) the value of
              the barrier at the mean switching current. (b) Normalized
              current versus normalized damping over ramp
              ratio~\cite{comm2}. (c) Mean energy divided by $k_BT$ of
              particles in the well (brown) and fraction of particles
              which have escaped (green) as a function of the applied
              current ($T=0.01$, $\gamma=10^{-5}$ and
              $\dot{I}=3.33\times 10^{-7}$).}
  \label{fig:isw}
\end{figure}

\textit{Switching current}--- In a typical JJ experiment the
probability distribution function of the junction switching current
$P(I)$ is measured performing many current-voltage curves where
current is continuously increased at a given rate. From these results
the mean switching current $I_{\rm sw}$ and its standard deviation can
be trivially computed. Such $P(I)$ can be easily related to the escape
rates $r(I)$ as~\cite{Fulton1974}
\begin{equation}
P(I)=r(I) \left( \frac{dI}{dt} \right)^{-1} \left( 1- \int_0^I P(u) du \right).
\label{eq:P(I)}
\end{equation}
Alike, escape rates can be computed from measured $P(I)$.

Figure~\ref{fig:isw}(a) shows our numerical results for the average
switching current and compares them to theoretical predictions. We
integrate Eq.~(\ref{langevin}) for an ensemble of thermalized
junctions.  Current is increased at a given ramp and switching events
are recorded. As expected BHL based predictions fail in the very low
damping regime. However, surprisingly, also DH is unable to explain
our numerical results, which lie in between both theories. This is due
to the competition between the equilibrium time of the system, given
by $\gamma^{-1}$, and the time order for the change of the current,
given by the inverse of the current ramp. Thus, switching in the very
low damping regime is a non-equilibrium process.  The coupling to the
external bath is so weak that other junctions are not able to reach
the thermal energy before switching. Thus, junctions escape in an {\em
  evaporative cooling} way where more energetic junctions switch first
and the ensemble is effectively cooled. This picture is confirmed in
Fig.~\ref{fig:isw}(c) where for a given damping we show the mean
energy for the trapped junctions as a function of the current and the
fraction of particles which have switched.

We also see in Fig.~\ref{fig:isw}(c) that particles escape with an
initial energy which goes from $E_{\rm in}= K_{\rm B}T$ to zero when
current is increased. The simplest way to introduce this fact in the
theory is to assume an average value of $E_{\rm in}= K_{\rm B}T/2$ and
use our Eq.~(\ref{mnz}). Figure~\ref{fig:isw}(a) shows that in this
way we are able to reproduce quite accurately the numerical
results. This correction turns out to be important when the average
barrier at the switching current is of the order of the thermal
energy. See that in the figure it is also plotted the value of the
barrier at the mean switching current (open symbols). Finally, using
Eqs.~(\ref{HTB-harmonic}),~(\ref{mnz}) and (\ref{eq:P(I)}) it can be
seen that in this region of the parameter space, the results depend on
the $\gamma/\dot I$ ratio, as confirmed in
Fig.~\ref{fig:isw}(b). Therefore our theory allows to estimate the
values for $\gamma/\dot I$ where non equilibrium corrections are
necessary.

Although presented in the framework of the JJ switching current
measurements, our results are further more general and apply to any
experiment where an activation rate is measured as a function of an
external parameter which can be controlled at will. An important issue
to study, it is the influence of the observed competition between two
different time scales on results for biased systems at higher values
of the damping and thus transfer our theoretical scheme from the
energy-diffusion regime to the phase-diffusion one. This is the
typical case for many of the current biological-physics
experiments~\cite{Hyeon2003,Dudko2009}

We thanks F. Falo and L.M. Flor\'{\i}a for discussions and critical
reading of the manuscript.  This work was supported by Spain MICINN
project FIS2008-01240, co-financed by FEDER funds.

\bibliography{lowd}

\begin{thebibliography}{25}
\bibitem[{com({\natexlab{b}})}]{comm2}
\bibinfo{note}{To normalize Eq.~(\ref{langevin}) we divide it by $V_0$ and time
  by $\omega^{-1}=\sqrt{m/V_0}$. Then , the adimensionalized parameters are
  $\tilde{\gamma}=\gamma/\omega=1.50 \gamma$, $\tilde{T}=k_B T / V_0=6.45 k_B
  T$ and $\widetilde{\dot{I}}=\dot{I} / (\omega V_0)=9.70 \dot{I}$.}
\end{thebibliography}

\end{document}